\documentstyle[12pt]{article}
\begin{document}
\begin{center}
   \vskip 2em
  {\LARGE BF gravity and the Immirzi parameter}
   \vskip 3em
  {\large R.  Capovilla${}^{(1)}$, M. Montesinos${}^{(1)}$, V. A.
Prieto${}^{(2)}$, and
E. Rojas${}^{(1)}$ \\[2em]}
\em{
 ${}^{(1)}$ Departamento de F\'{\i}sica \\
 Centro de Investigaci\'on
y de Estudios Avanzados del I.P.N. \\
Apdo Postal 14-740, 07000 M\'exico,
D. F.,
MEXICO \\[1em]
${}^{(2)}$ Department of Physics\\
 University  of Maryland  \\
 College Park MD 20742-4111, USA }
\end{center}
 \vskip 1em
\begin{abstract}
We propose a novel  BF-type formulation of real four-dimensional gravity,
which generalizes previous models. In particular, it allows for an arbitrary
Immirzi parameter. We also construct the analogue of the Urbantke metric for
this model.
\end{abstract}
\date{\today}
\vskip 3em
Real general relativity can be formulated as a constrained
first-order BF-type theory of the form \cite{CDJ1}
(for an earlier alternative approach, see \cite{Plebanski})
\begin{equation}
S [B, A , \phi, \mu ]  = \int \left[ B^{IJ} \wedge F_{IJ} (A) +
G(B , \phi, \mu ) \right] \, ,
\label{eq:action}
\end{equation}
where $B^{IJ} = - B^{JI} $ are six real 2-forms,
$A^{IJ} = - A^{JI}$ is an $SO(3,1)$ connection, with $F^{IJ}
= d A^{IJ} + A^I{}_K \wedge A^{KJ}$ its curvature
(Lorentz indices $I,J, \dots = 0,1,2,3$ are raised and
lowered with the Minkowski metric $\eta_{IJ}$).
$G(B , \phi, \mu )$ denotes a constraint quadratic in the
2-forms $B^{IJ}$. Its role is to implement that,
for some
tetrad $e^I$,  the 2-forms take the form
$B^{IJ} = \ast (e^I \wedge e^J) $,
with $\ast$ the duality operator on Lorentz indices (
$\ast^2 = -1$).
When the constraint is solved, substitution back in the
action of this specific form would then recover general
relativity in its first-order tetrad formulation.
For Euclidean gravity, we have
$\eta^{IJ} \to \delta^{IJ}$, the connection is valued in $SO(4)$,
and $\ast^2 = + 1 $.

The constraint is of the form
\begin{equation}
G ( B , \phi , \mu ) =
- {1 \over 2} \phi_{IJKL} B^{IJ} \wedge B^{KL}
+ \mu H (\phi )\,,
\label{eq:constraint}
\end{equation}
with $\phi_{IJKL}$  a  Lagrange multiplier with obvious symmetries $
\phi_{IJKL} = - \phi_{JIKL}= -\phi_{IJLK} = \phi_{KLIJ}$. It has 21
independent components. Since this is one too many, as the $B^{IJ}$ have 36
and the $e^I$ have 16, one needs to impose some extra conditions on $\phi$,
via the 4-form Lagrange multiplier $\mu$, which sets $H (\phi) = 0$. The
following conditions were considered \cite{CDJ1,Rei1,FdP}:
\begin{eqnarray}
H_1 &=& \phi_{IJ}{}^{IJ} = 0\,, \label{eq:h1} \\
H_2 &=& \phi_{IJKL} \varepsilon^{IJKL} = 0 \label{eq:h2} \,.
\end{eqnarray}
The solution of the associated constraints on the $B^{IJ}$ in terms of an
arbitrary real tetrad $e^I$ leads to a modified version of tetrad gravity,
\begin{equation}
S [ e , A ] = \alpha \; \int  \ast  ( e^I \wedge e^J ) \wedge F_{IJ} (A)  +
\beta  \; \int e^I \wedge e^J  \wedge F_{IJ} (A) \,.
\label{eq:tetrad}
\end{equation}
where for $H_1$, we have the four independent solutions $\alpha = \pm 1$ and
$\beta = \pm 1$, whereas from $H_2$ one obtains the four independent
solutions $\alpha = \pm 1, \beta = 0$, or $\alpha = 0 ,\beta = \pm 1$. Apart
from the annoying sign ambiguities, the first term gives the Einstein-Hilbert
action, and, at least for non-degenerate tetrads, the second term vanishes on
shell.

These formulations have been put to use both in the Lorentzian and the
Euclidean cases in the context of various real four-dimensional approaches to
quantum gravity that have come to be known as spin foam models \cite{spin},
or Feynman diagrams for gravity \cite{RRdP}. These approaches were motivated
initially by canonical quantum gravity, first in its complex version based on
the Ashtekar phase space variables \cite{QGcomplex}, then in its real version
based on the Ashtekar-Barbero phase space variables \cite{QGreal}. In the
latter there is an arbitrary real parameter, which enters in the spectra of
geometric operators, which has come to be known as the Immirzi parameter
\cite{Giorgio}. It corresponds to ${\alpha}/{\beta}$ in (\ref{eq:tetrad}).
One would expect it to play a role also in these various four-dimensional
approaches, but as their continuum analog is given by the BF models described
above, this appears  not to be the case.

The purpose of this note is to point out that the Immirzi parameter emerges
naturally if one considers a condition more general than (\ref{eq:h1}) and
(\ref{eq:h2}). This seems to have been overlooked in previous investigations.
(This possibility was advocated first in \cite{Violeta}, but at the time its
relevance was not recognized.)  We assume that, rather than vanishing
separately, the invariants are proportional so that
\begin{equation}
H_3 = a_1 \phi_{IJ}{}^{IJ} + a_2 \phi_{IJKL} \varepsilon^{IJKL}
= 0 \,,
\label{eq:h3}
\end{equation}
with $a_1, a_2$ arbitrary constants.
The previous cases are obtained as either one vanishes.
We emphasize that this possibility is available only in
four dimensions. In the BF formulation of higher dimensional
gravity, there is no difference between (\ref{eq:h1})
and the equivalent of (\ref{eq:h2}) \cite{BFhd}.

Variation of the action with respect to the Lagrange multipliers,
and taking the appropriate traces, gives that the constraints on the
$B^{IJ}$ are
\begin{eqnarray}
B^{IJ}\wedge B^{KL} & = & \frac{1}{6} (B^{MN}\wedge B_{MN})
\eta^{[I\mid K \mid}
\eta^{J ] L} + \frac{\epsilon}{12} ( B^{MN} \wedge {\ast B}_{MN})
\varepsilon^{IJKL} \,, \label{eq:const1} \\
2 a_2 B^{IJ} \wedge B_{IJ} &=& \epsilon  a_1 B^{IJ} \wedge
\ast B_{IJ} \,, \label{eq:const2}
\end{eqnarray}
with $\epsilon =1$ in the Euclidean case and $\epsilon =- 1$ in the
Lorentzian one. It is easy to show that in this case the 2-forms are given
in terms  of some tetrad $e^I$  by
\begin{equation}
B^{IJ} =  \alpha \ast ( e^I \wedge e^J ) +  \beta e^I \wedge e^J \,,
\label{eq:sol}
\end{equation}
with $\alpha , \beta$ arbitrary non-vanishing real parameters that satisfy
$\alpha^2 \neq \beta^2$ (the case $\alpha^2 = \beta^2$ gives the special case
(\ref{eq:h1})). It is only a matter of calculation to prove that this
condition is necessary. For the sufficiency it is convenient to break
explicit internal Lorentz invariance and express the constraints in terms of
$B^{0i}, B^{ij}$ ($i$, $j =$ 1,2,3), and the problem reduces to a set of
constraints on three 2-forms, say $B^{0i}$, with (\ref{eq:sol}) as its
immediate solution. The ratio $ \beta / \alpha$ is determined algebraically
by the ratio $a_2 / a_1 = (\alpha^2 + \epsilon \beta^2 )/ 4 \alpha \beta$, as
follows from (\ref{eq:const2}). Substitution of (\ref{eq:sol}) in
(\ref{eq:action}) gives the tetrad action (\ref{eq:tetrad}), with arbitrary
Immirzi parameter ${\alpha}/{\beta}$.

When either $a_1$ or $a_2$ vanishes, we see by inspection of
(\ref{eq:const2}) that we obtain the degenerate cases in which $B^{IJ} \wedge
B_{IJ}$ or $B^{IJ} \wedge \ast B_{IJ}$ vanish respectively. This modifies
(\ref{eq:const1}), and the form of the $B^{IJ}$ is restricted accordingly.

It is interesting to construct the metric directly in terms of the $B^{IJ}$.
Guided by its analogue in the self-dual case, which has come to be known
as the Urbantke metric \cite{Urbantke,CDJM}, we
consider the two possible invariants, of weight two, cubic in the $B^{IJ}$,
\begin{eqnarray}
{\cal G}^{\mu\nu} &=& \tilde{B}^{\mu\alpha}{}^{IJ} \;
B_{\alpha\beta}{}^{KL}\;
\tilde{B}^{\beta\nu}{}^{MN} \; \eta_{IN} \;\epsilon_{JKLM} \,, \\
 {\cal H}^{\mu\nu} &=& \tilde{B}^{\mu\alpha}{}^{IJ} \;
B_{\alpha\beta}{}^{KL}\;
\tilde{B}^{\beta\nu}{}^{MN} \; \eta_{IN}\; \eta_{JK} \;\eta_{LM}\,,
 \end{eqnarray}
where $ \tilde{B}^{\mu\nu}{}^{IJ} = \varepsilon^{\mu\nu\rho\sigma}
\; B_{\rho\sigma}{}^{IJ} $,  $
 \varepsilon^{\mu\nu\rho\sigma} $ denotes the spacetime Levi-Civita tensor
density, and greek indices denote spacetime indices.

Using (\ref{eq:sol}) the following is obtained
\begin{eqnarray}
{\cal G}^{\mu\nu} &=&
6 \alpha \; (\epsilon \alpha^2 + 3 \beta^2 )
\; g \; g^{\mu\nu} \,, \\
 {\cal H}^{\mu\nu} &=&
3 \beta \; (\epsilon \beta^2 + 3 \alpha^2 )
g \;  g^{\mu\nu} \; \,
 \end{eqnarray}
where $g^{\mu\nu} = e^\mu{}_I e^\nu{}_J \eta^{IJ} $, $g$ its determinant and
$e^\mu{}_I$ the inverse tetrad. These expressions allow us to express the
(densitized) metric in various ways in terms of the $B^{IJ}$.  In the special
cases of $H_1$ and $H_2$ is enough one of the metrics ${\cal G}^{\mu\nu}$ or
${\cal H}^{\mu\nu}$. Considering that we are assuming $\alpha^2 \neq
\beta^2$,  the  natural choice is
\begin{equation}
g g^{\mu\nu} = {\epsilon \over 3 (\beta^4 - \alpha^4 )}
\left( \beta {\cal H}^{\mu\nu} - {1 \over 2}
\alpha {\cal G}^{\mu\nu} \right)\,.
\end{equation}

The Hamiltonian formulation of the action (\ref{eq:tetrad}) has been
performed in \cite{Holst} for arbitrary parameters, and leads to the
Ashtekar-Barbero phase space variables \cite{AB} (see also
\cite{Alexandrov}). Alternatively, one can perform the canonical analysis
directly from the BF action (\ref{eq:action}), with the condition
(\ref{eq:h3}), arriving at the same result \cite{Violeta}.

We emphasize that the constraint (\ref{eq:constraint})
is what distinguishes a topological field theory
(the BF part), from a theory with local degrees of
freedom like gravity. It is crucial to understand
how to impose it at the quantum level. In particular, our observation
suggests that in current discrete models which use a triangulation of
spacetime, the partition function should be extended to
include non-simple representations associated with
the triangles, and change appropriately the intertwiners.
One expects that under this generalization, one would
arrive at the same conclusion of the canonical approaches:
a one-parameter ambiguity in the spectra of geometric
operators such as area and volume.

\vskip 1cm
\noindent{\bf Acknowledgements}
\vskip .5cm
RC and ER acknowledge partial support from CONACyT grant 32187-E.
We thank J Zapata for useful criticism of an early draft of the paper.

\end{document}